\documentclass[twocolumn,showpacs,preprintnumbers,amsmath,amssymb]{revtex4-1}
\usepackage{graphicx}
\usepackage{textcomp}

\begin{document}
%\draft
\title{Time-reversed optical parametric oscillation}
  \normalsize
\author{Stefano Longhi %\footnote{Author's email address: longhi@fisi.polimi.it}
}
\address{Dipartimento di Fisica, Politecnico di Milano, Piazza L. da Vinci
32, I-20133 Milano, Italy}

%\date{.}
%
\bigskip
\begin{abstract}
\noindent In recent works, the idea of time-reversed laser
oscillation has been proposed and demonstrated to realize a
two-channel coherent perfect absorber [Y.D. Chong et al., Phys. Rev.
Lett. {\bf 105},
 053901 (2010); W.Wan et al., Science {\bf 331}, 889 (2011)]. Here the time
 reversal of optical parametric oscillation in a nonlinear
$\chi^{(2)}$ medium is considered and shown to realize a coherent
perfect absorber for colored incident signal and idler fields. A
detailed analysis is presented for the time-reversed process of
mirrorless optical parametric oscillation in the full nonlinear
regime.
\end{abstract}


% 42.65.Yj    Optical parametric oscillators and amplifiers
% 42.65.Ky    Frequency conversion; harmonic generation, including higher-order harmonic generation
% 42.65.Sf    Dynamics of nonlinear optical systems; optical instabilities, optical chaos and complexity, and optical spatio-temporal dynamics

\maketitle

{\it Introduction.} Time-reversal symmetry plays an important role
in classical electromagnetic theory and nonrelativistic quantum
mechanics \cite{timereversal}. Such a symmetry implies that, if a
particular physical process is allowed, then there also exists a
"time-reversed process", which is obtained by changing the sign of
the time variable in the dynamical equations. In recent works, the
time-reversed process of laser oscillation has been explored, both
theoretically \cite{StonePRL2010} and experimentally
\cite{StoneScience2011}, and shown to realize a coherent perfect
absorber (CPA). Under additional symmetry constraints, the processes
of lasing and CPA can even occur simultaneously, and a CPA-laser
device can be in principle realized
\cite{LonghiPRA2010,StonePRL2011}. Since in a steady-state process
time reversal corresponds to interchanging incoming and outgoing
fields, the time-reversed process of a laser instability corresponds
to perfect absorption of incoming light fields. Such a phenomenon is
rather general and expected to occur for the time reversal of other
optical instabilities. Among these, optical parametric oscillation
(OPO) provides an important example of instability in nonlinear
optics, which is at the core of the operation of such important
devices as optical parametric oscillators \cite{P0,P1,P2,P2bis}. In
a parametric oscillator at the onset of instability, a pump field at
frequency $\omega_3$ propagating in a nonlinear $\chi^{(2)}$ crystal
generates down-converted fields at frequencies $\omega_1$ (the
signal field) and $\omega_2=\omega_3-\omega_1$ (the idler field),
provided that the phase matching (momentum conservation) condition
is satisfied. Generally, at least one of the signal or idler waves
resonates in an optical cavity. In the time-reversed process,
exactly the opposite occurs: two coherent signal and idler fields,
with appropriate amplitude and phase relationship incident onto the
pumped nonlinear crystal, are completely annihilated, thus realizing
a kind of "colored" CPA device. While a CPA based on time reversal
of a laser usually converts the absorbed light into heat
\cite{StonePRL2010,StoneScience2011}, in the time reversal of OPO
the annihilated photons are converted into light of different
frequency (the sum-frequency field). Moreover, since the OPO
instability and its time-reversal process are allowed in the same
medium, an ordinary OPO device can behave at the instability like a
CPA-laser device \cite{LonghiPRA2010,StonePRL2011}, i.e. it can
simultaneously emit outgoing colored fields and absorb incoming
colored fields with appropriate phase relationships. This important
property shows that a conventional OPO system at threshold is a
CPA-laser of a certain kind. In this Letter we consider specifically
the time-reversed process of OPO for counterpropagating photons,
i.e. the time-reversed process of backward OPO \cite{P2,P3,P4},
which enables a simple analytical study. Backward OPO describes an
interesting three-wave interaction process in a quadratic medium
which has the unique property of automatically establishing
distributed feedback, thus allowing mirrorless OPO. Mirrorless OPO
was theoretically predicted in the early days of nonlinear optics
\cite{P3}, however its experimental demonstration has been reported
only in recent years using short-period quasi-phase-matching (QPM)
nonlinear crystals \cite{P6,P7}. Backward parametric interaction is
also useful for the generation of entangled twin photons \cite{ent}.
\par {\it Time-reversal of OPO
instability: general aspects.} We consider interaction of three
quasi-monochromatic waves at carrier frequencies $\omega_1$ (signal
field), $\omega_2$ (idler field) and $\omega_3=\omega_1+\omega_2$
(pump field) in a nonlinear $\chi^{(2)}$ medium in the framework of
 classical nonlinear optics equations. In the scalar
approximation, the electric field $\mathcal{E}(\mathbf{r},t)$
satisfies the wave equation $\nabla^2_{\mathbf{r}}
\mathcal{E}-(1/c_0^2) \partial^2_t \mathcal{E}=(1/\epsilon_0 c_0^2)
\partial^2_t \mathcal{P}$, where $c_0$ is the speed of light in
vacuum, $\epsilon_0$ is the vacuum permittivity,
$\nabla^2_{\mathbf{r}}$ is the spatial Laplacian, and
$\mathcal{P}=\mathcal{P}_L+\mathcal{P}_{NL}$ is the polarization of
the medium, which includes the linear ($\mathcal{P}_L$) and
nonlinear ($\mathcal{P}_{NL}$) terms. For nonlinear interaction in a
quadratic medium with QPM, one can assume $\mathcal{P}_{NL}=
\epsilon_0 \chi^{(2)}\mathcal{E}^2$, where $\chi^{(2)}(\mathbf{r}) $
is the spatially-modulated second-order susceptibility of the
medium. After setting
$\mathcal{E}(\mathbf{r},t)=(1/2)[E_1(\mathbf{r},t) \exp(i \omega_1
t)+E_2(\mathbf{r},t) \exp(i \omega_2 t)+ E_3(\mathbf{r},t) \exp(i
\omega_3 t)+c.c.]$, where $E_l(\mathbf{r},t)$ varies slowly with
respect to time $t$ over one oscillation cycle $2 \pi/ \omega_l$
($l=1,2,3$), neglecting group velocity dispersion and indicating by
$\epsilon_l(\mathbf{r})$ the relative dielectric constant of the
medium at the frequency $\omega_l$, the envelopes $E_l$ satisfy the
coupled wave equations (see, for instance, \cite{Longhi02})
\begin{eqnarray}
\left( \nabla^2_{\mathbf{r}}+ \frac{\omega_{1,2}^2
\epsilon_{1,2}}{c_0^2}-\frac{2ik_{1,2}}{v_{1,2}}
\partial_t \right) E_{1,2} & = & - \frac{\omega_{1,2}^2}{c_0^2}
 \chi^{(2)}E_{2,1}^*E_3  \;\;   \nonumber \\
 \left( \nabla^2_{\mathbf{r}}+ \frac{\omega_3^2
\epsilon_3}{c_0^2}-\frac{2ik_3}{v_3}
\partial_t \right) E_3 & = & - \frac{\omega_3^2}{c_0^2}
 \chi^{(2)}E_1E_2  \;\;\;\;\;\;\;
\end{eqnarray}
where $k_l= \omega_l n_l/c_0$, $n_l$ and $v_l$ $(l=1,2,3$) are the
wave number, bulk refractive index and group velocity at frequency
$\omega_l$, respectively. For a lossless medium, the dielectric
constant $\epsilon_l(\mathbf{r})$ is real-valued and can be
generally written as $\epsilon_l(\mathbf{r})=n_l^2+2 n_l \Delta
n_l(\mathbf{r})$, where $\Delta n_l(\mathbf{r})$ is a small
correction to the bulk refractive index $n_l$ that accounts for e.g.
index guiding or cavity effects for the signal and/or idler fields,
like in singly- or doubly-resonant OPOs. Equations (1) should be
supplemented with the boundary conditions at the crystal (or cavity)
boundaries. In an OPO geometry, a traveling-wave pump field shines
on the crystal, and the signal and idler fields are spontaneously
generated at the OPO instability as stationary outgoing waves. Time
reversal of the underlying equations implies that, if
$E_l(\mathbf{r})$ is a stationary solution to Eqs.(1) at the OPO
instability point, then $E_l^* (\mathbf{r})$ is still a solution to
the same equations. Since the complex conjugation replaces outgoing
waves with incoming waves, the time-reversed process of OPO is the
perfect annihilation of the incoming signal and idler fields, i.e. a
colored CPA is realized. A schematic of the OPO instability and of
its time-reversal is depicted in Fig.1. It should be noted that the
present analysis is strictly valid only for classical fields, and
phenomena involving few photons as well as quantum noise are not
considered in such a semiclassical approach. If quantum fields are
considered, spontaneous parametric down-conversion occurs, and the
complete annihilation (conversion to the pump field) of both
incident fields cannot be strictly reached.

\begin{figure}
\includegraphics[scale=0.42]{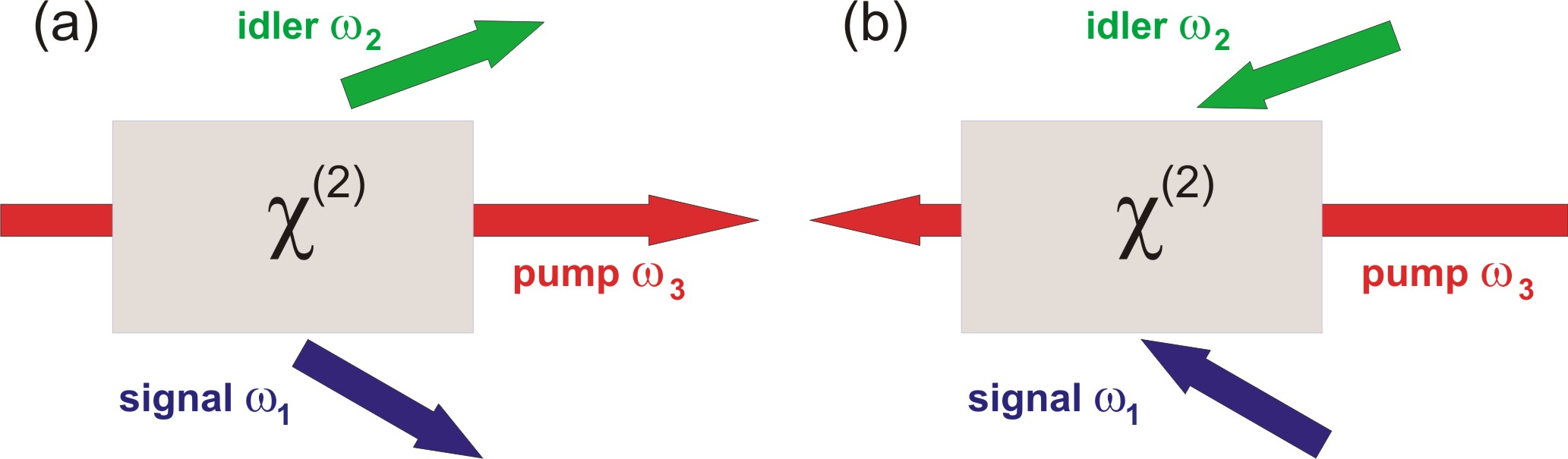}
\caption{(Color online) Schematic of (a) an OPO instability, and (b)
its time-reversed process, corresponding to a colored CPA.}
\end{figure}
\par
{\it Time-reversed backward OPO.} In the following, we will focus
our analysis to the backward OPO configuration in a QPM crystal of
length $L$, which allows for mirrorless OPO [i.e. $\Delta n_l=0$ in
Eqs.(1)]. In the plane-wave approximation \cite{P4}, for a backward
configuration the pump and signal fields copropagate along the
positive $z$ direction, whereas the idler wave is
counter-propagating along the negative $z$ direction, as shown in
Fig.2(a). Phase matching of the interaction is ensured by a QPM
grating $\chi^{(2)}(z+\Lambda)=\chi^{(2)}(z)$ with period $\Lambda$
satisfying the first-order QPM condition $2 \pi /
\Lambda=|k_3+k_2-k_1|$. After setting $E_{1,3}(z,t)=(1/ \kappa l)
(\omega_{1,3}/n_{1,3})^{1/2} a_{1,3}(x,t) \exp(-ik_{1,3}z)$,
$E_{2}(z,t)=(1/ \kappa l) (\omega_{2}/n_{2})^{1/2} a_2(z,t)
\exp(ik_{2}z)$ for the forward- and backward-propagating waves and
introducing the normalized spatial propagation distance $Z=z/L$,
from Eqs.(1) the following coupled-mode equations can be obtained
for the dimensionless envelopes $a_l$ after averaging the
rapidly-oscillating terms in $z$ \cite{P4,Longhi02}
\begin{eqnarray}
\pm \partial_Z a_{1,2}+(L/v_{1,2}) \partial_t a_{1,2} & = & -i
a_{2,1}^{*} a_3 \nonumber \\
\partial_Z a_3 +(L/v_3) \partial_t a_3 & = & -i a_1 a_2
\end{eqnarray}
where $\kappa=(1/2c_0)(\omega_1 \omega_2 \omega_3/ n_1 n_2
n_3)^{1/2} \chi_M $ and $\chi_M $ is the Fourier amplitude of
$\chi^{(2)}(z)$ at the spatial frequency $2 \pi / \Lambda$. Without
loss of generality, $\kappa$ is assumed to be real-valued and
positive. The intensities $I_l$ of the three waves are given by
$I_l= \epsilon_0 c_0 n_1n_2n_3 \lambda_1 \lambda_2 \lambda_3
|a_l|^2/[2 \pi^2 L^2 \chi_M^{2} n_l \lambda_l]$, where $\lambda_l=2
\pi c_0/ \omega_l$ are the (vacuum) wavelengths of the three waves
($l=1,2,3$). To investigate the time reversal of OPO, we assume that
a continuous-wave pump with amplitude $a_3(0) \equiv a_{30}$ is
injected into the nonlinear crystal, together with two signal and
idler waves with amplitudes
$a_1(0)$ and $a_2(1)$, respectively.\\
\begin{figure}
\includegraphics[scale=0.6]{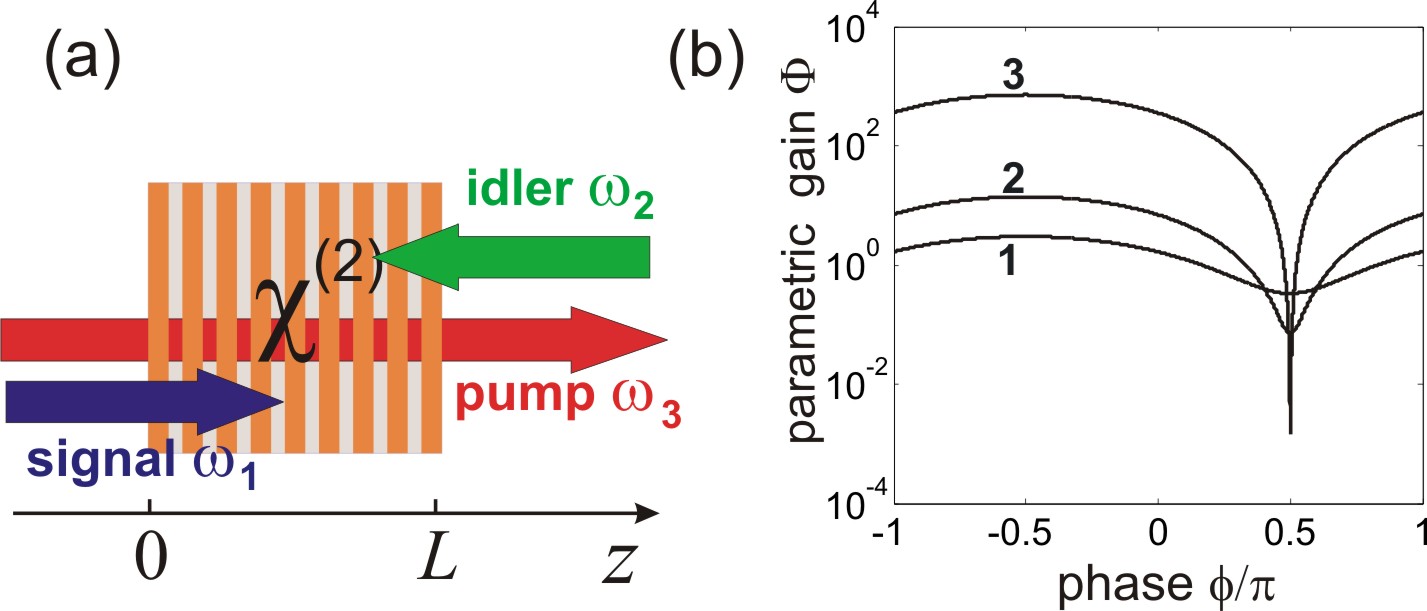}
\caption{(Color online) (a) Schematic of backward parametric
interaction of pump, signal and idler waves in a nonlinear QPM
crystal. (b) Behavior of the parametric gain $\Phi$ versus the phase
difference $\phi$ of incident signal and idler fields for increasing
values of the normalized input pump amplitude $a_{30}$ in the
no-pump-depletion approximation and below the OPO threshold. Curve
(1): $a_{30}=\pi/6$; curve (2): $a_{30}=\pi/3$; curve (3):
$a_{30}=\pi/2.1$. The dip at $\phi=\pi/2$ is the signature of
colored CPA.}
\end{figure}
Let us first assume that the amplitudes of injected signal and idler
fields are much smaller than the pump amplitude $a_{30}$
(small-signal analysis). Then in the no-pump-depletion
approximation, $a_3(Z) \simeq a_{30}$, Eqs.(2) reduce to a system of
two coupled linear equations for the amplitudes $a_1(Z)$ and
$a_2^{*}(Z)$ \cite{P2}, which yield the algebraic relation
$(a_1(1),a_2^*(1))^T=\mathcal{M}(a_1(0),a_2^*(0))^T$ between the
amplitudes of signal and idler fields at input ($Z=0$) and output
($Z=1$) crystal planes. Assuming without loss of generality a
positive and real-valued amplitude $a_{30}$, the coefficients of the
transfer matrix $\mathcal{M}$ read explicitly
$\mathcal{M}_{11}=\mathcal{M}_{22}=\cos(a_{30})$ and
$\mathcal{M}_{12}=\mathcal{M}_{21}=-i \sin(a_{30})$. The vanishing
of $\mathcal{M}_{22}$ as the pump amplitude is increased from zero
defines the mirrorless OPO threshold, which is attained at
$a_{30}=\pi/2$ (see, e.g.,  \cite{P2,P4}). It is worth observing
that the transfer matrix $\mathcal{M}$ has the same structure as
that encountered in the $\mathcal{PT}$-symmetric CPA-laser device,
recently proposed in Refs.\cite{LonghiPRA2010,StonePRL2011}. In
particular, since $\mathcal{M}_{11}=\mathcal{M}_{22}$, the
simultaneous vanishing of $\mathcal{M}_{11}$ and $\mathcal{M}_{22}$
at the OPO instability threshold indicates that the system can
simultaneously emits coherent signal and idler waves and acts as a
colored CPA for appropriate excitation. This means that an OPO {\em
at the instability threshold} basically realizes a kind of colored
CPA-laser system. Like in the CPA-laser device, colored CPA requires
coherent excitation of the crystal with signal and idler fields of
appropriate amplitude and phase relationship, and the degree of
absorption/amplification can be conveniently tuned by varying the
relative phase of the two incoming beams. In fact, let us consider a
pump level close to (but below) the threshold for OPO instability,
and let us assume that the crystal is simultaneously illuminated,
from opposite sides, by signal and idler waves with the same
amplitude $A$ and phase difference $\phi$, namely let us assume
$a_1(0)=A \exp(i \phi)$ and $a_2^*(1)=A$. The parametric gain for
signal and idler waves after propagation in the crystal is given by
$\Phi(\phi)=|a_1(1)/A|^2=|a_2^*(0)/A|^2=\{\cos^2(\phi)+[\sin(\phi)-\sin(a_{30})]^2\}/
\cos^2(a_{30})$. A typical behavior of $\Phi(\phi)$ versus the phase
difference $\phi$ is shown in Fig.2(b) for increasing values of the
pump level $a_{30}$. Note that, at $\phi= \pi/2$ a dip is observed
in the gain curve, with $\Phi(\pi/2) \rightarrow 0$ as the threshold
of OPO is approached. Such a dip is the clear signature of CPA. Far
from $\phi=\pi/2$, the system behaves conversely like a parametric
amplifier ($\Phi>1$), with a parametric gain that diverges as the
OPO threshold is approached. The curves shown in Fig.2(b) clearly
indicate that an optical parametric amplifier is capable of
attenuating strongly as well when coherently illuminated from both
sides, for an appropriate relative phase of the incoming fields.
Such a result basically stems from the well-known phase-sensitivity
of the frequency conversion process, which controls the flow of
energy from the pump to the signal and idler waves.
\par
The previous considerations are based on a small-signal analysis of
Eqs.(2), which neglects pump depletion effects. However, a more
comprehensive analysis should consider the full nonlinear regime,
i.e. should account for pump depletion effects. Such an analysis is
of major importance to check the stability of the CPA regime against
the spontaneous emission of outgoing signal and idler waves. The
exact solution corresponding to CPA in the nonlinear regime can be
constructed following a similar procedure used in the theory of
backward OPO \cite{P4}, but with modified boundary conditions. Let
us search for a stationary solution to Eqs.(2) of the form
\begin{equation}
\bar{a}_1(Z)=a_0^* \sin [\theta(Z)] \; , \;\; \bar{a}_2(Z)=-i a_0
\cos [\theta(Z)],
\end{equation}
where $a_0$ is a complex parameter and the angle $\theta(Z)$ is an
unknown function. Substitution of Eq.(3) into Eqs.(2) yields the
following coupled equations for $\bar{a}_3(Z)$ and $\theta(Z)$
\begin{equation}
\partial_Z \theta=\bar{a}_3 \; , \;\; \partial_Z
\bar{a}_3=-(|a_0|^2/2) \sin(2 \theta).
\end{equation}
The colored CPA regime is attained by imposing the boundary
conditions $\bar{a}_1(1)=\bar{a}_2(0)=0$ for the signal and idler
fields, and $\bar{a}_3(0)=a_{30}$ for the pump field. The former
conditions can be satisfied by taking $\theta(0)=\pi/2$ and
$\theta(1)=n \pi $, where $n=1,2,3,...$. Note that in this case
$|a_0|$ corresponds to the normalized amplitudes of the
counter-propagating signal and idler fields incident onto the
crystal at the $Z=0$ and $Z=1$ facets, respectively. For an assigned
 pump level $a_{30}$, the value of $|a_0|$ has to be consistently
 determined from Eqs.(4) with the appropriate boundary conditions.
 In fact, from Eqs.(4) it follows that the angle $\theta(Z)$
 satisfies the pendulum equation $(d^2 \theta/dZ^2)+(|a_0|^2/2) \sin (2
 \theta)=0$, which admits of the constant of motion (energy) $E=(1/2) (d
 \theta/dZ)^2-(|a_0|^2/4) \cos(2 \theta)$. Hence the following
 relation holds
 \begin{equation}
\int_{\theta(0)=\pi/2}^{\theta(1)= n \pi} \frac{d
\theta}{\sqrt{2E+(|a_0|^2/2) \cos(2 \theta)}}= \int_0^1 dZ=1.
 \end{equation}
The value of the energy $E$ is determined by the boundary condition
at $Z=0$ and reads $E=(1/2)a_{30}^2+|a_0|^2/4$. After setting $\eta
\equiv |a_0/a_{30}|^2$, from Eq.(5) one readily obtains
\begin{equation}
\left( \sqrt{1+\eta} \right) a_{30}=(2n-1) K \left(
\sqrt{\frac{\eta}{1+\eta}} \right)
\end{equation}
where $K(\xi) \equiv \int_0^{\pi/2} dq/[1-\xi^2 \sin^2 (q)]^{1/2}$
is the complete elliptic integral of the first kind. For a given
order $n$, Eq.(6) implicitly defines the relation between $|a_0|$
and $a_{30}$, which is depicted in Fig.3(a) for the lowest-order
($n=1$) state. Hence, to realize the CPA condition, two conditions
should be satisfied: (i)the relative phase between signal and
conjugate idler waves is fixed according to Eq.(3); (ii) for a given
injected pump level $a_{30}$, the normalized signal and idler wave
amplitudes injected into the crystal from the two sides should take
the same value $|a_0|$ defined by the curve of Fig.3(a). In
particular, for small values of $|a_0|$ the amplitude $a_{30}$ of
the pump wave that realizes the CPA condition is exactly the
threshold value $\pi/2$ of OPO instability, according to the
small-signal analysis previously discussed. The explicit behaviors
of $a_l(Z)$ along the crystal are then obtained by solving Eqs.(4)
and using Eqs.(3). Note that the phase of the amplitude $a_0$, which
defines the absolute phases of the injected signal and idler fields
, remains fully undetermined, so that if $\{\bar{a}_1(Z),
\bar{a}_2(Z), \bar{a}_3(Z)\}$ is a stationary solution to Eqs.(2)
corresponding to CPA, then $\{\bar{a}_1(Z) \exp(i \varphi),
\bar{a}_2(Z) \exp(-i \varphi), \bar{a}_3(Z)\}$ is a solution as well
for an arbitrary phase $\varphi$. As an example, Fig.3(b) shows the
numerically-computed behavior of the normalized field intensities
$|\bar{a}_l(Z)|^2$ for pump, signal and idler fields along the
crystal for the lowest-order CPA solution ($n=1$) corresponding to
$\eta=5$ [point C of Fig.3(a)]. Note that the annihilation of the
signal and idler photons corresponds to the creation of pump
photons, rather than to the depletion as in the backward OPO
\cite{P4}. Higher-order ($n \geq 2$) CPA solutions show oscillating
power exchange among signal, idler and pump fields along the
crystal. As an example, Fig.3(c) shows the behavior of
$|\bar{a}_l(Z)|^2$ along the crystal for the $n=2$ mode
corresponding to $\eta=5$.
\par
The stability of the CPA solution can be investigated by a standard
linear stability analysis \cite{supp}. After setting
$a_l(Z,t)=\bar{a}(Z)+\delta a_l(Z,t)$ in Eqs.(2), where $\delta
a_l(Z,t)$ are small perturbations ($l=1,2,3$), the most general
solution to the linearized equations $\delta a_l(Z,t)$ is given by
an arbitrary superposition of solutions of the form $\delta
a_{l}(Z,t)=u_l(Z) \exp(\lambda t)+v_l^*(Z) \exp(\lambda^*t)$, where
the vector of perturbation amplitudes $\mathbf{v} \equiv
(u_1,u_2,u_3,v_1,v_2,v_3)^T$ satisfies the linear system
$(d\mathbf{v}/dZ)=\mathcal{G}(Z,\lambda) \mathbf{v}$
 and $\lambda$ is
a complex parameter whose real part determines the growth or decay
rate of the perturbation. The explicit form of the $6 \times 6$
matrix $\mathcal{G}(Z,\lambda)$ is given in the supplementary
material \cite{supp}. The eigenvalues $\lambda$ are then determined
after imposing the boundary conditions for the perturbations at the
$Z=0$ and $Z=1$ planes \cite{supp}.
\begin{figure}
\includegraphics[scale=0.42]{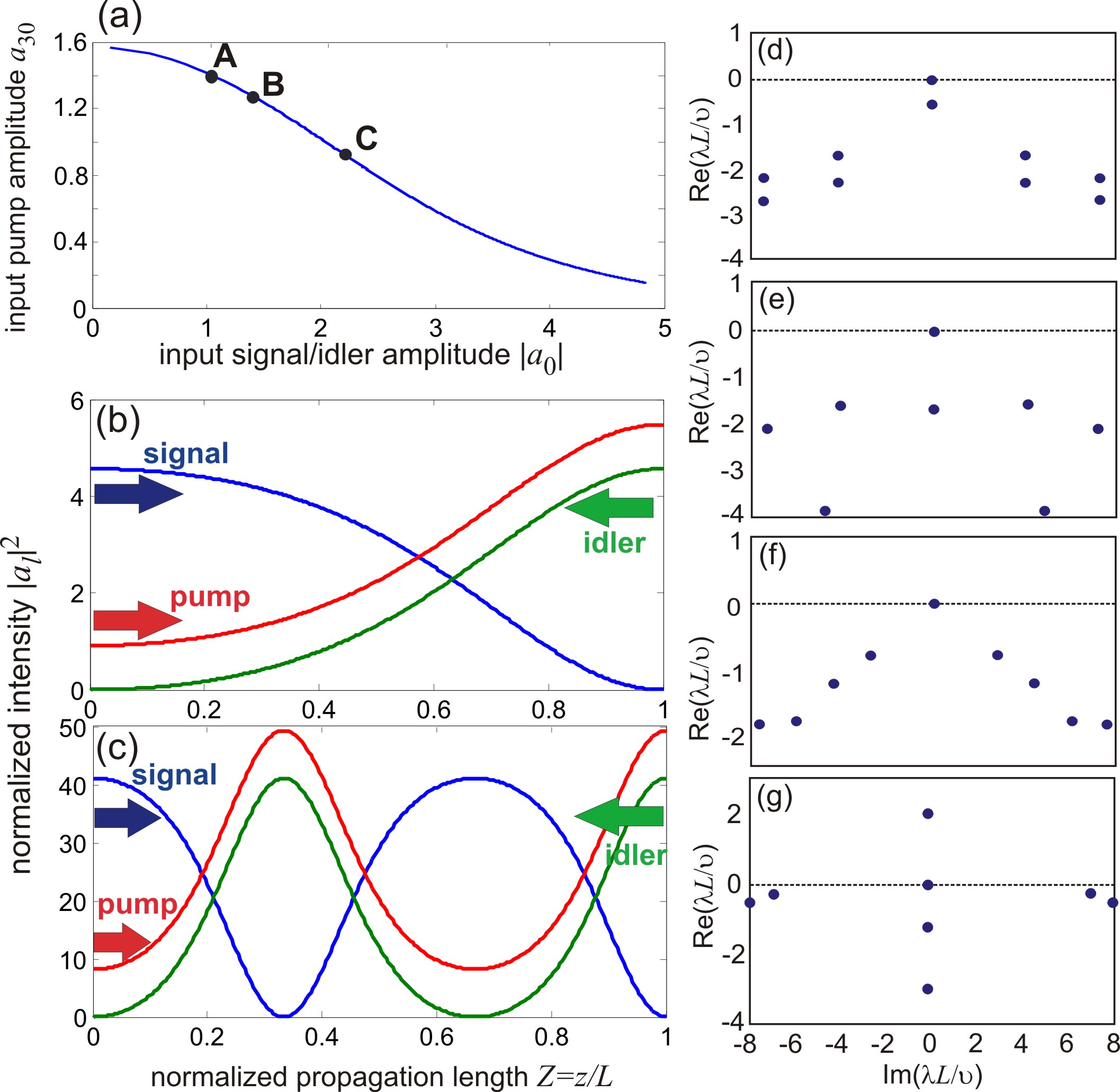}
\caption{(Color online) (a) Behavior of the input pump amplitude
$a_{30}$ versus the amplitude $|a_0|$ of input signal and idler
fields corresponding to the lowest-order CPA solution, numerically
computed from Eq.(6) with $n=1$. The points A, B and C in the figure
correspond to $\eta=0.5$, $\eta=1$ and $\eta=5$, respectively, where
$\eta \equiv |a_0/a_{30}|^2$. (b) Behavior of the normalized
intensities $|a_l(Z)|^2$ of pump, signal and idler fields inside the
nonlinear crystal for the lowest-order ($n=1$) CPA solution
corresponding to $\eta=5$ [point C of Fig.3(a)]. (c) Same as (b),
but for the second-order ($n=2$) CPA solution with $\eta=5$. (d-f)
Loci of the eigenvalues $\lambda$ of the linearized equations for
the lowest-order CPA solution corresponding to points A, B and C of
Fig.3(a). (g) Same as (d-f), but for the second-order CPA solution
of Fig.3(c). In the computation of the eigenvalues, the same group
velocity $v_1=v_2=v_3=v$ has been assumed for the three waves.}
\end{figure}
Owing to the phase invariance of the CPA solution, a neutrally
stable mode with eigenvalue $\lambda=0$ is always found. An
instability arises provided that the real part of at least one of
the other eigenvalues is positive. Extended numerical simulations
indicate that the lowest-order ($n=1$) CPA solution is stable at any
pump level, whereas higher-order ($n \geq 2$) CPA solutions are
linearly unstable. As an example, Figs.3(d-f) show the
numerically-computed loci of the eigenvalues $\lambda$ for the
lowest-order ($n=1$) CPA solution at $\eta=0.5$, $\eta=1$ and
$\eta=5$, respectively, corresponding to points A, B and C in
Fig.3(a). Note that all the eigenvalues lie in the lower-half
complex plane, and thus there are not unstable modes. Conversely,
Fig.3(g) depicts the eigenvalues for the $n=2$ CPA solution at
$\eta=5$, indicating the existence of an unstable mode associated to
the eignevalue with a positive real part.

\par {\it Conclusion.}
In this work, time reversal of optical parametric oscillation has
been proposed and shown to realize a kind of "colored" coherent
perfect absorber. A detailed analysis has been presented for the
time reversal of mirrorless OPO. At threshold and in the
no-pump-depletion limit, it has been shown that the OPO can
simultaneously emit outgoing coherent signal and idler waves, and
perfectly absorb incoming coherent signal and idler waves, thus
realizing a kind of CPA-laser device
\cite{LonghiPRA2010,StonePRL2011}. As compared to previous studies
on CPA and CPA-laser systems
\cite{StonePRL2010,StoneScience2011,LonghiPRA2010,StonePRL2011}, CPA
operation has been extended here into the nonlinear regime. In
particular,  for the mirrorless OPO a proof of the stability of the
colored CPA operation for the fundamental CPA mode has been
presented.

%\par
%This work was supported by the Italian MIUR (Grant No.
%PRIN-20082YCAAK, "Analogie ottico-quantistiche in strutture
%fotoniche a guida d'onda").

\end{document}